\documentclass[10pt,english]{article}
\usepackage{lmodern}

\usepackage[T1]{fontenc}
\usepackage[latin9]{inputenc}
\usepackage{geometry}
\usepackage{babel}
\usepackage{array}
\usepackage{float}
\usepackage{multirow}
\usepackage{amsmath}
\usepackage{amssymb}
\usepackage{graphicx}
\usepackage{esint}
\usepackage{subscript}
\usepackage{float}
\usepackage{color}
\usepackage[bookmarks=false,linkcolor=red]{hyperref}
\hypersetup{colorlinks=true,breaklinks,urlcolor=[rgb]{0,0,1},citecolor=[rgb]{0,0.6,0}}
\usepackage{url}
\usepackage{multirow} 
\usepackage{array}
\usepackage[font=footnotesize,format=plain,labelfont=bf, margin=0.8cm]{caption}
\usepackage{multirow}
\usepackage{array}
\usepackage{booktabs}
\usepackage{rotating}
\usepackage[numbers,sort&compress]{natbib}

\begin{document}

\title{On evaluation of ShARP passive rainfall retrievals over snow-covered
land surfaces and coastal zones}

\author{Ardeshir M. Ebtehaj\textsuperscript{1}, Rafael L. Bras\textsuperscript{1}, and Efi Foufoula-Georgiou\textsuperscript{2}}
\maketitle
{\footnotesize{}\textsuperscript{1}School of Civil and Environmental
Engineering, Georgia Institute of Technology, Atlanta, Georgia, USA.}{\footnotesize \par}

{\footnotesize{}\textsuperscript{2}Department of Civil, Environmental, and Geo-Engineering,
University of Minnesota, Minneapolis, Minnesota, USA.}{\footnotesize \par}

\begin{abstract}
For precipitation retrievals over land, using satellite measurements
in microwave bands, it is important to properly discriminate the weak
rainfall signals from strong and highly variable background surface
emission. Traditionally, land rainfall retrieval methods often rely
on a weak signal of rainfall scattering on high-frequency channels
(85 GHz) and make use of empirical thresholding and regression-based
techniques. Due to the increased ground surface signal interference,
precipitation retrieval over radiometrically complex land surfaces---especially
over snow-covered lands, deserts and coastal areas---is of
particular challenge for this class of retrieval techniques. This
paper evaluates the results by the recently proposed Shrunken locally
linear embedding Algorithm for Retrieval of Precipitation (ShARP),
over a radiometrically complex terrain and coastal areas using the
data provided by the Tropical Rainfall Measuring Mission (TRMM) satellite.
To this end, the ShARP retrieval experiments are performed over a
region in South and Southeast Asia, partly covering the Tibetan Highlands,
Himalayas, Ganges-Brahmaputra-Meghna river basins and its delta. This
region is unique in terms of its diverse land surface radiation regime
and precipitation type within the TRMM field of view. We elucidate
promising results by ShARP over snow covered land surfaces and at
the vicinity of coastlines, in comparison with the land rainfall retrievals
of the standard TRMM-2A12 product. Specifically, using the TRMM-2A25
radar product as a reference, we provide evidence that the ShARP algorithm
can significantly reduce the rainfall over estimation due to the background
snow contamination and markedly improve detection and retrieval of
rainfall at the vicinity of coastlines. During the calendar year 2013,
we demonstrate that over the study domain the root mean squared difference
can be reduced up to 38\% annually, while the reduction can reach
up to 70\% during the cold months.
\end{abstract}

\section{Introduction}

Accurate estimation of rainfall from space is of paramount importance
for hydrologic and land-atmosphere studies, especially where ground
measurements are not readily available. The launch of the Tropical
Rainfall Measuring (TRMM) satellite set a benchmark for spaceborne
estimation of rainfall over the tropics ($40^{\circ}$S-N), providing
an invaluable set of decadal rainfall observations with unprecedented
accuracy and coverage. The recently launched Global Precipitation
Measurement (GPM) core satellite, together with a constellation of
partner satellites, will provide a considerably extended spatial coverage
of precipitation over higher latitudes ($68^{\circ}$S-N). The TRMM
satellite carries aboard a Microwave Imager (TMI), which is an orthogonally
polarized radiometer with 9 channels centered on frequencies ranging
from 10.7-to-85.5 GHz \citep{Kum98} with a swath width of 878 km
(after 2001, during the post boost-era). Aboard the GPM core satellite,
the GPM Microwave Imager (GMI) provides 13 dual-polarized channels
with central frequencies within 10.65-183.3 GHz, which provides more
detailed information about the frozen hydrometeors \citep{Hou2013}
than its predecessor over a swath width of 904 km. On board both of
these two satellites there are precipitation radars providing high-resolution
measurements of precipitation reflectivity--concurrent with the radiometric
measurements within a narrower swath width. The TRMM precipitation
radar (PR) is a single-frequency Ku band radar (13 GHz) with a 247
km swath width (during the post boost-era), while the GPM carries
a dual-frequency radar (DPR) operating at Ku and Ka bands (13 and
35 GHz) with a swath width of 245 km.

In the microwave frequency bands ranging from 1-200 GHz, it is well
understood that the transparent window channels contain a mixture
of surface and precipitation signatures in the upwelling radiative
fluxes reaching the top of the atmosphere \citep{Grody88}. Over ocean,
rainfall emission contrasts itself well from the radiometrically uniform,
cold and polarized background emission. This signature can be properly
explained via the absorption-emission laws of the atmospheric radiative
transfer, giving rise to well-developed physically based algorithms
for rainfall retrieval over ocean \citep[among others]{Wil77,WilCc91,BerC92,Changetal_99,Olson89,Mugnai1993,KumG94a,KumG94b,Smith_et_al_94,Petty_94a,Petty_94b,Bauer01,KumOWG96,Kumetal01,KumRCRB10}.
Among these, the so-called Bayesian algorithms have provided a viable
path for operational rainfall retrievals. This family of algorithms
relies on a priori collected, statistically representative, database
that contains hydrometer profiles and their measured or simulated
spectral radiances at the top of the atmosphere. Measuring radiometric
fluxes from space, candidate spectral radiances are found in the database
and then their corresponding rainfall profiles are used to estimate/retrieve
precipitation values of interest. Among these, the \citet{WilCc91}
and \citet{Kumetal01} algorithms have been successfully deployed
for operational ocean retrieval of the Advanced Microwave Scanning
Radiometer (AMSR-E) on board the Aqua satellite and TMI. Conversely,
over land, radiometrically warm background emission typically masks
the microwave rainfall emission signal, causing the retrieval algorithms
to often hinge upon weak rainfall scattering on high frequency channels.
It turns out that, as the wavelength approaches the particle size,
the scattering due to the large raindrops and/or ice crystals causes
depression in the field of brightness temperatures (BT) at high frequency
channels (e.g., 85 GHz). Traditionally, the radiometrically cold raining
areas are discerned from the non-raining background via a fixed thresholding
scheme and then related to the surface rainfall through empirical
regression-based equations. Among these are the seminal 85 GHz scattering
index (SI) by \citet{Gro91} used for the SSM/I sensor; the one suggested
by \citet{Ferraro1994} and \citet{FerM95} partly used for the SSM/I
and the AMSR-E sensors \citet{WilKF03}; and the used $\text{SI}={\rm Tb_{22v}}-{\rm Tb_{85v}}$
for the operational TMI-2A12 product \citep{Kumetal01,Gop10}. A thorough
review of the SI methods can be found in \citep{Seto05}. 

Specifically, the TMI land retrieval algorithm makes use of the scattering
index $\text{SI}={\rm Tb_{22v}}-{\rm Tb_{85v}}$ greater than \textasciitilde{}8
Kelvin to discriminate the rainfall signal from the non-raining background,
as the water vapor channel of 22 GHz is not very sensitive to the
precipitation scattering. Although the scattering-based approaches
have been the cornerstone of overland microwave rainfall retrievals,
they often suffer from some well-known drawbacks including: 1) The
SI-based screening naturally cannot properly detect warm precipitation
regimes lacking scattering effects of the ice crystals \citep{LiuZ09}.
2) This technique often confuses the high-frequency depressions over
snow-covered lands or deserts with those caused by ice/rainfall scattering
and produces biased rainfall estimates. 3) Due to the underlying complexity
and transient structure of the background radiation regime, this class
of methods often misses light rainfall at the vicinity of coastlines.
In the GPM era, these drawbacks are of great importance, especially
for hydrologic applications in temperate and cold climate regimes
with complex land-surface radiation dynamics.

It turns out that microwave emissions in the frequency range 1-37
GHz respond well to the heterogeneity of the land surface feature--e.g.,
soil texture, roughness, moisture content; vegetation density, pattern,
water content; snow cover and water equivalent--while also contain
partial information about the atmospheric constituents. For bare to
sparsely vegetated land, soil layers with higher moisture content
are less emissive and more polarized in low frequencies, especially
in 1-37 GHz \citep{Njoku1996,NjokuLi99}. Vegetation and rough surfaces
often reduce the polarization signature due to their Lambertian-like
reflection properties and are typically more emissive than their bare
soil counterparts. The Desert surfaces contain little moisture and
exhibit large diurnal variability and polarization in their thermal
emission, especially for high frequencies with small penetration depth.
Moreover, desert surfaces scatter the upwelling radiation in a manner
very similar to light precipitation \citep{GroW08dessert}. Owing
to the snow metamorphism, snow emissivity exhibits a markedly varying
dynamics as a function of snow grain size, density, depth, water content
and ice crust effects \citep{Grody2008}. Typically, snow emissivity
sharply decreases from low to high microwave frequencies, especially
for fresh snow, giving rise to appreciably cold radiometric temperatures
in high frequency channels (i.e., $\geq$37 GHz), very similar to
the scattering signatures of ice crystals in raining clouds, making
the retrieval process notoriously ill-posed. 

Therefore, it is likely that non-raining and raining scenes with different
atmospheric constituents, exhibit similar scattering signals at the
top of the atmosphere (TOA), especially over the aforementioned complex
surfaces. This equifinality or non-uniqueness is naturally the main
source of ambiguity for land rainfall retrievals, especially when
the method only exploits the information content of a few high-frequency
channels. Hence, we need retrieval methods that can account for various
interactions among different channels rather than only using a few
high-frequency bands. In other words, all of the available spectral
channels that contain partial information about the land surface emissivity
and hydrometeor profile need to be robustly exploited and properly
encoded in a multispectral sense to improve rainfall retrieval over
land and better mitigate its ill-posedness. For instance, over snow-covered
surfaces, the low-frequency channels (10-37 GHz) are radiometrically
colder than their bare or vegetated counterpart surfaces (see, \citet{Grody2008}
and Figure \ref{fig:1}). Therefore, this piece of information needs
to be properly accounted for to help deciphering whether the depressions
in high-frequency channels are due to the rainfall scattering or perhaps
low emissivity of snow-covered land surfaces. 

To this end, a growing body of research has been recently directed
toward a better understanding of land surface emissivity dynamics
for improving the rainfall retrieval over land \citep[e.g.,][among others]{SkofB11,Ferraro_et_al_13,Ringerud14,Turk2013a,TurkHad14}.
These efforts are of central importance in the GPM era, especially
as the attempts are more toward Bayesian approaches for over land
retrievals \citep[e.g.,][]{KumRCRB10,MunchakGail_2013}. Microwave
land surface emissivity is typically estimated by purely physical
modeling and/or radiative energy balance using satellite measurements.
Physical modeling requires a Land Surface Model (LSM) coupled with
a Radiative Transfer Model (RTM) to capture the space-time variability
of both the structural and electromagnetic properties of the land
surfaces. This family of methods \citep[e.g.,][]{Ringerud14} requires
very detailed information about geophysical parameters controlling
the land surface-atmosphere interactions, soil-vegetation dynamics
and their interactions with subsurface hydrologic processes. The availability
of such parameters is often limited \citep{Prigent06} at a global
scale, especially at the fine space-time resolution of interest for
rainfall retrieval. Energy balance approaches typically make use of
satellite measurements of spectral BTs and estimates of the surface
skin and atmospheric temperature-moisture profiles. These estimates
are typically used to close the radiation budget at TOA to approximate
the \textquotedblleft clear sky\textquotedblright{} emissivity values
\citep[among others]{Prigent97,Prigent06,Moncet2011,Norouz11}, which
may be used to improve rainfall retrieval over land \citep{Turk2013a,TurkHad14}. 

In this paper, we do not directly attempt to estimate the land surface
spectral emissivity and explicitly incorporate it in the rainfall
retrieval algorithm. Rather, we explore the idea of \textquotedblleft implicitly\textquotedblright{}
encoding the land surface information content across all available
frequency channels, as part of the recently proposed Bayesian ShARP
methodology by \citet{EbtBF15}, briefly reviewed in Section \ref{sec:2}.
In essence, the a priori database in this Bayesian algorithm is properly
organized in an algebraically tractable manner via two fat matrices,
called \textquotedblleft rainfall\textquotedblright{} and \textquotedblleft spectral
dictionaries\textquotedblright . Spectral dictionary contains a large
a priori collection of BT measurements, while the rainfall dictionary
encompasses their corresponding simulated or observed rainfall profiles.
Given a pixel-level measurement of spectral BT values, this algorithm
relies on a nearest neighbor search that uses information content
of all frequency channels to properly narrow down the retrieval problem
to a few physically relevant spectral candidates and their rainfall
profiles. In this step, our algorithm classifies the measured BT values
into raining and non-raining signatures via a simple probabilistic
voting rule (detection step). Then for the raining BT measurements,
it exploits a modern regularized weighted least-squares estimator
to retrieve the rainfall values of interest (estimation step). Section
\ref{sec:3} is devoted to presenting some promising rainfall retrievals
over a region in South and Southeast Asia with diverse land-surface
conditions and precipitation regimes. This study area covers the Ganges-Brahmaputra-Meghna
basin and its delta while encompasses some important land surface
features including the Tibetan highlands, Himalayan range, Thar Dessert,
Hengduan Mountains, and the Malabar coasts in the western Indian subcontinent
(Figure \ref{fig:1}). Due to the snow covered and frozen ground,
especially over the Tibetan plateau and Hengduan Mountains, this region
has been notoriously problematic for the SI-based microwave retrievals
\citep{Wangetal09}. Moreover, due to frequent inundation of the Ganges
delta, dense irrigated agriculture, complex earth surface dynamics
of the river mouths and presence of near shore orographic features,
this region is naturally a suitable test bed for examining the quality
of rainfall retrievals over wet surfaces and coastal zones. It is
important to note that, although ShARP can be applied uniformly for
retrievals both over ocean and land, in this paper, we confine our
consideration only to its quantitative evaluation over land. To this
end, we populate the spectral and rainfall dictionaries using collocated
TMI-1B11 and PR-2A25 observations and compare the retrieval results
with the standard TMI-2A12 (version 7) operational land rainfall retrievals.
We draw some concluding remarks and envision future research directions
in Section \ref{sec:4}.

\section{Retrieval Methodology\label{sec:2}}

\subsection{Shrunken Locally Linear Embedding for Retrieval of Precipitation
(ShARP)}

In this subsection we provide a brief explanation of ShARP, while
a detailed account can be found in \citep{EbtBF15}. It is known that
the passive rainfall retrieval in microwave bands can be defined as
a mathematical inverse problem. In this inverse problem, we aim to
obtain the rainfall profile (input) from the measured spectral BTs
at TOA (output) while the BTs can be related to the rainfall profile
through the physical laws of radiative transfer. Let us assume that,
at a pixel-level, the observed spectral BTs at $n_{c}$ channels are
denoted by an $n_{c}$-element vector $\mathbf{y}=\left(y_{1},\, y_{2},\,\ldots,\, y_{n_{c}}\right)^{T}\in\mathfrak{R}^{n_{c}}$,
while the rainfall profile of interest, sampled at $n_{r}$ levels
through the atmospheric depth is $\mathbf{x}=\left(x_{1},\, x_{2},\,\ldots,\, x_{n_{r}}\right)^{T}\in\mathfrak{R}^{n_{r}}$.
We can then assume that BTs are related to the rainfall profile via
the following nonlinear mapping

\begin{equation}
\mathbf{y}=\mathcal{F}\left(\mathbf{x}\right)+\mathbf{v},
\end{equation}

where $\mathcal{F}:\,\mathbf{x}\rightarrow\mathbf{y}$ can be considered
as a functional representation of a forward RTM and $\mathbf{v}\in\mathfrak{R}^{n_{c}}$
collectively represents the model/measurement error. As is evident,
this mapping is extremely nonlinear, especially over land where the
rainfall signal is severely masked by the background earth emission.
At the same time, as we only measure BT values within a few spectral
bands, different rainfall profiles may give rise to identical or very
similar spectral BTs making this inverse problem severely ill-posed. 

Let us assume that the a priori collected database is denoted by $\mathcal{D}=\left\{ \left(\mathbf{b}_{i},\,\mathbf{r}_{i}\right)\right\} _{i=1}^{M}$,
where the pairs $\mathbf{b}_{i}=\left\{ b_{1i},\, b_{2i},\,\ldots,\, b_{n_{c}i}\right\} ^{T}\in\mathfrak{R}^{n_{c}}$and
$\mathbf{r}_{i}=\left\{ r_{1i},\, r_{2i},\,\ldots,\, r_{n_{r}i}\right\} ^{T}\in\mathfrak{R}^{n_{r}}$
are $n_{c}$- and $n_{r}$-element vectors representing pixel-level
spectral BTs and their measured/simulated rainfall profiles. For notational
convenience, we organize these pairs in two matrices $\mathbf{B}=\left[\mathbf{b}_{1}|\ldots|\mathbf{b}_{M}\right]^{T}\in\mathfrak{R}^{n_{c}\times M}$
and $\mathbf{R}=\left[\mathbf{r}_{1}|\ldots|\mathbf{r}_{M}\right]^{T}\in\mathfrak{R}^{n_{r}\times M}$,
called spectral and rainfall dictionaries, respectively.

Let us assume that $\mathbf{y}\in\mathfrak{R}^{n_{c}}$denotes a pixel-level
satellite measurement of the spectral BTs. In the first step, our
algorithm isolates, in the Euclidean sense, a set $\mathcal{S}$ of
``spectral neighbors'' as the $k$-nearest neighbors of $\mathbf{y}$
in the column space of $\mathbf{B}$. Then these spectral neighbors
and their corresponding rainfall profiles in $\mathbf{R}$ are stored
in the pair of spectral $\mathbf{B}_{\mathcal{S}}\in\mathfrak{R}^{n_{c}\times k}$
and rainfall $\mathbf{R}_{\mathcal{S}}\in\mathfrak{R}^{n_{r}\times k}$
sub-dictionaries, respectively. In this step (detection step), we
label $\mathbf{y}$ as a raining or non-raining pixel, using a simple
classification rule. Specifically, we count the number of profiles
in $\mathbf{R}_{\mathcal{S}}$ which are raining at the surface. If
the number of raining profiles is greater than or equal to $pk$,
then we label $\mathbf{y}$ as raining and attempt to estimate its
rainfall profile. Note that, here, $p\in\left(0,\,1\right]$ is a
probability measure that controls the probability of hit and false
alarm. In words, a smaller $p$ value gives rise to larger probability
of hit while a larger value reduces the probability of false alarm.
We typically set $p=0.5$, which corresponds to a majority vote rule
for rain/no-rain classification. In the next step (estimation step),
the algorithm attempts to approximate the observed raining spectral
BTs, using a linear combination of the neighboring spectral candidates
in $\mathbf{B}_{\mathcal{S}}$ as follows:

\begin{equation}
\begin{aligned} & \underset{\mathbf{c}}{\text{minimize}} &  & \left\Vert \mathbf{W}^{1/2}\left(\mathbf{y}-\mathbf{B}_{\mathcal{S}}\mathbf{c}\right)\right\Vert _{2}^{2}+\lambda_{1}\left\Vert \mathbf{c}\right\Vert _{1}+\lambda_{2}\left\Vert \mathbf{c}\right\Vert _{2}^{2}\\
 & \text{subject to} &  & \mathbf{c}\succeq0,\enskip\mathbf{1}^{{\rm T}}\mathbf{c}=1,
\end{aligned}
\label{eq:2}
\end{equation}

\noindent where, $\ell_{p}$-norm is $\left\Vert \mathbf{c}\right\Vert _{p}^{p}=\sum_{i}c_{i}^{p}$,
$\mathbf{c}\mathbf{\succeq}0$ denotes element-wise non-negativity,
$\lambda_{1}$ and $\lambda_{2}$ are positive regularization parameters,
$\mathbf{W}\in\mathfrak{R}^{n_{c}\times n_{c}}$ is a positive definite
weight matrix that determines the relative importance of each channel
and $\mathbf{1}^{T}=\left[1,\ldots,1\right]^{T}\in\mathfrak{R}^{k}$.
This weight matrix encodes rainfall signal-to-noise ratio in each
channel. For instance, it can be properly designed to give more weights
to lower (e.g., 10 GHz) or higher frequencies (e.g., 85 GHz) while
raining BT measurements are over ocean or land, respectively. Note
that, the non-negativity constraint is required to be physically consistent
with the positivity of the BTs in degrees Kelvin, while the sum to
one constraint assures that the estimates are locally unbiased. Obtaining
the representation coefficients by solving (\ref{eq:2}), the rainfall
profile is then retrieved as $\hat{\mathbf{x}}=\mathbf{R}_{\mathcal{S}}\mathbf{\hat{c}}$.
Here, we consider a convex combination of the regularization terms
by assuming  $\lambda_{1}=\lambda\left(1-\alpha\right)$ and $\lambda_{2}=\lambda\alpha$,
where $\alpha\in\left(0,\,1\right]$ \citep{ZouH05}.

\subsection{TMI-2A12 and differences with ShARP}

As briefly noted, the version 7 of the standard TMI-2A12 product uses
a scattering index together with regression-based models for rainfall
retrieval over land. However, over ocean, this product relies on the
Goddard Profiling Algorithm \citep[GPROF: see,][]{Kumetal01,Kummerow2007,KumRCRB10},
which is called a Bayesian algorithm as it makes use of a priori collected
database of rainfall-radiance pairs.

\subsubsection{TMI-2A12: land retrieval}

For each raining pixel, the algorithm interpolates between stratiform
and convective rainfall mechanisms. To this end, it attempts to estimate
a convective probability indicator $p_{c}$ \citep[see,][]{McCF03}
and then obtains an estimation of the near-surface rain rate as follows:

\begin{equation}
\hat{\mathbf{x}}=p_{c}\mathbf{\hat{x}}_{\text{conv}}+\left(1-p_{c}\right)\hat{\mathbf{x}}_{\text{strat}}
\end{equation}

\noindent where $\mathbf{\hat{x}}_{\text{conv}}$ and $\mathbf{\hat{x}}_{\text{strat}}$
are rainfall estimates for convective and stratiform conditions as, 

\begin{equation}
\hat{\mathbf{x}}_{\text{conv}}=-11.77\text{e-}6\,{\rm Tb}_{{\rm 85v}}^{3}+8.027\text{e-}3\,{\rm Tb}_{{\rm 85v}}^{2}-1.946\,{\rm Tb}_{{\rm 85v}}+182.68
\end{equation}

\noindent and 

\begin{equation}
\hat{\mathbf{x}}_{\text{strat}}=-70.8\text{e-}3\,{\rm Tb_{85v}}+19.7.
\end{equation}

The above regression equations are derived using collocated TMI-PR
datasets explained in \citep{Liu08}. For a full exposition to the
version 7 status of the land algorithm one can refer to \citet{Gop10}.

\subsubsection{TMI-2A12: ocean retrieval}

Over ocean the TMI-2A12 relies on the GPROF algorithm. As of noted
by \citet{Kummerow2007}, in this algorithm the expected value of
rainfall profile of interest is obtained as follows:

\begin{equation}
\mathbf{\hat{x}}=\mathbb{E}\left(\mathbf{x}\right)=\frac{\sum_{i}\mathbf{r}_{i}\exp\left\{ -0.5\left[\mathbf{y}-\mathcal{F}\left(\mathbf{r}_{i}\right)\right]^{T}\mathbf{V}^{-1}\left[\mathbf{y}-\mathcal{F}\left(\mathbf{r}_{i}\right)\right]\right\} }{\sum_{i}\exp\left\{ -0.5\left[\mathbf{y}-\mathcal{F}\left(\mathbf{r}_{i}\right)\right]^{T}\mathbf{V}^{-1}\left[\mathbf{y}-\mathcal{F}\left(\mathbf{r}_{i}\right)\right]\right\} },\label{eq:6}
\end{equation}

\noindent where $\mathbf{V}$ is the covariance of model/observation error and
$\left\{ \mathbf{r}_{i}\right\} _{i=1}^{M}$ denotes the derived rainfall
profiles from a set of existing cloud-resolving model (CRM) simulations
for which, the associated vector of brightness temperatures $\left\{ \mathbf{b}_{i}\right\} _{i=1}^{M}$
are obtained through an RTM as $\mathbf{b}_{i}=\mathcal{F}\left(\mathbf{r}_{i}\right)$.
As is evident, the estimator in (\ref{eq:6}) uses the inverse distance
Gaussian weighting function to lineally combine all of the rainfall
profiles in the database. It is important to note that although the
exponential function decays very quickly but it never goes to zero. As
a result, ``all rainfall profiles (predictors)'' in the database are linearly combined
in this estimator. Note that, in version 7, the algorithm uses a CRM generated
database that is made consistent with the TRMM-PR observations as
well \citep{KumRCRB10}. 

As is evident, the ShARP algorithm uses a unified Bayesian approach
over land and ocean and shares some similarities with the GPROF algorithm,
while exhibits notable distinctions. Indeed, ShARP has its heritage
in GPROF algorithm in the sense that it obtains some optimal representation
coefficients in the spectral space and then uses them to combine the
their corresponding rainfall profiles. However, there are two main
technical distinctions between the two algorithms. First, GPROF uses
\textquotedblleft all model simulated profiles in the atmosphere\textendash radiative
model database\textquotedblright , while ShARP only focuses dynamically
on a few isolated ``spectral neighbors'' in the database, which
are closest to the satellite measurement in the Euclidean sense. This
step in ShARP allows us to control the probability of detection and
false alarm important for over land retrievals. We will see later
on that this step not only allows us to screen out physically irrelevant
candidates, but also provides regionalization skill, which can make
the rainfall retrievals consistent with their underlying land surfaces.
Second, GPROF uses a basic inverse distance Gaussian weighting interpolation
scheme by \citet{Olson1996} to linearly combine all of the rainfall
profiles in the database, which assumes uniform prior probability
for the rainfall profiles \citep[see, equation 3 in][]{Kummerow2007}.
However, ShARP relies on a modern variational method in (\ref{eq:2}),
which can be interpreted as a Bayesian estimator \citep{EbtBF15} with a mixture of Gaussian and Laplace distributions as the prior density.
This estimator obtains the coefficients in a weighted least-squares
sense and promotes sparsity in the solution by assigning non-zero
coefficients only to a few relevant ``spectral neighbors''. By leveraging
sparsity, this estimator confines itself only to a few very relevant
spectral neighbors, which can lead to improved recovery of high-intensity
rainfall and prevent overly smooth retrievals. Moreover, due to the
variational nature of this estimator, this framework can be easily
extended for optimal integration of multiple sets of databases, models
and measurements across different satellites/instruments.

\section{Results\label{sec:3}}

To form the ShARP dictionaries, one may use the results of physical
modeling, analogous to the GPROF over ocean or just focus on satellite
observations (e.g., TMI and PR) as presented in many previous efforts
\citep[among others]{GreA02,Skofronick-Jackson2003,Greetal04,PetL13a,PetL13b}.
In this paper, we only focus on demonstrating the advantages of ShARP
over land and use the collocated TMI-1B11 (version 7) products and
surface PR-2A25 to populate the dictionaries.

\subsection{Study area}

\begin{figure}[H]
\noindent \begin{centering}
\includegraphics[width=0.70\paperwidth]{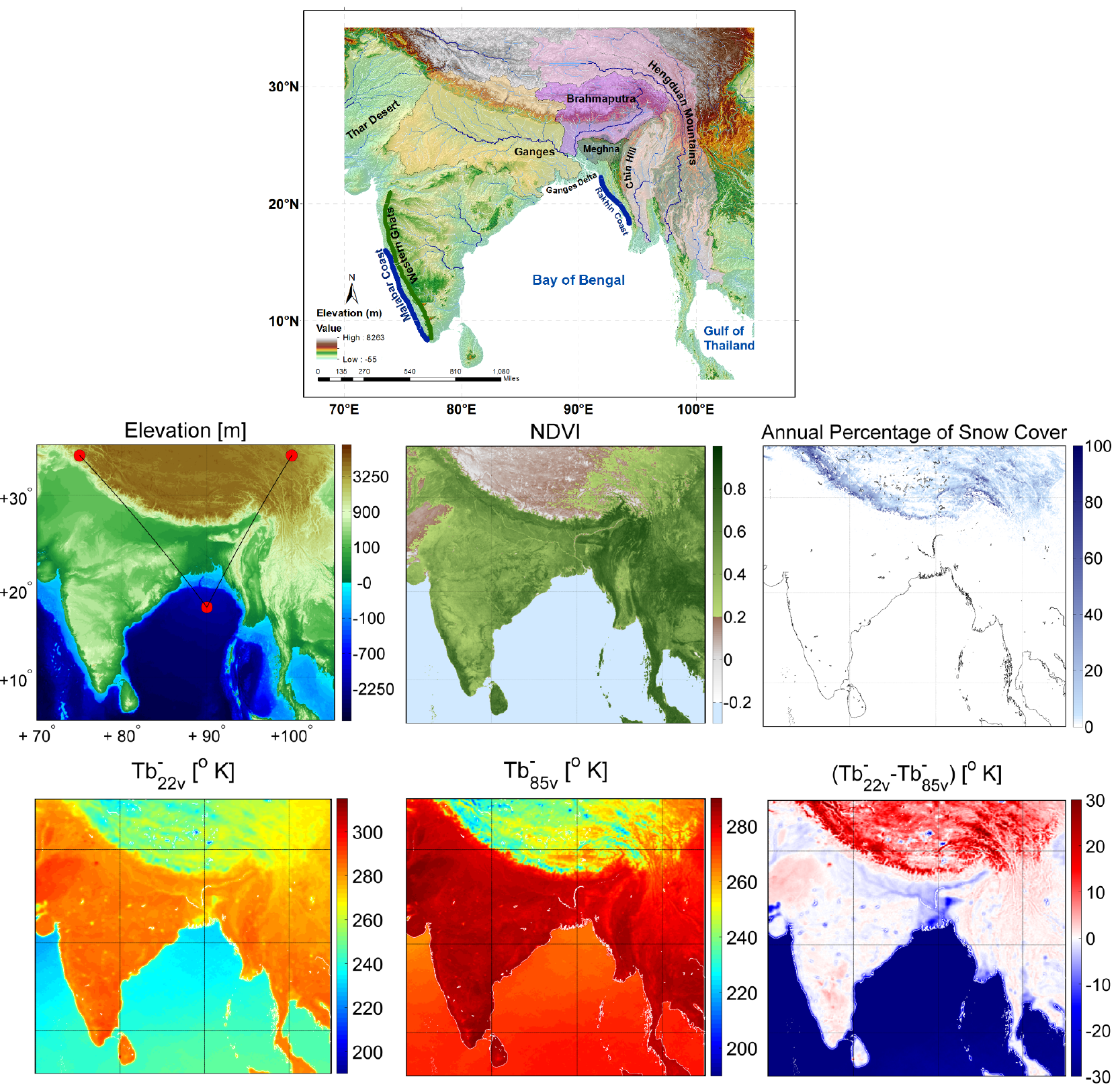}
\par\end{centering}

\protect\caption{Geographic boundaries of the study region (top panel) and some of
its land-surface characteristics. Middle row panels: the digital elevation
model, annual Normalized Difference Vegetation Index (NDVI) and percentage
of snow covered lands in 2013. Bottom row panels: The annual average
of the free-rain scene brightness temperatures measured by the TMI
sensor at frequency channels 22v, 85v and their difference (i.e.,
${\rm Tb_{22v}^{-}}-{\rm Tb_{85v}^{-}}$). For the shown tracks in
the digital elevation map, transects of all TMI channels, their corresponding
NDVI and snow percentage are shown in Figure \ref{fig:2}.\label{fig:1}}
\end{figure}

The study region (Figure \ref{fig:1}) is confined between 70-105E
longitudes and 5-35N latitudes, extending from the Indus river basin
and Thar Desert along the border of Pakistan and India to the Hengduan
Mountains in southwest China and Rakhin state at the western Mianmar.
From north to south the region covers from the Tibetan highlands to
the northern coastlines of the Bay of Bengal. This region is unique
in terms of its diverse climate regimes, biodiversity, landforms,
precipitation patterns and types. In the northwest, the region contains
the arid Thar Desert, while in the southwest it covers the coastlines
of the Indian Peninsula with a tropical wet climate influenced by
the Western Ghats orographic features. In the north, the elevated
Tibetan plateau in high altitudes has cold arid steppe climate. The
plateau is a glacier landform with a large number of brackish isolated
lakes and is often covered with snow in cold seasons, especially at
the vicinity of the Himalayan foothills. The Himalayan range sharply
divides the cold upstream climate from the temperate climate of downstream
lowlands at the base and has permanent snowpack in high elevations.
It is important to note that the study area encompasses the Ganges-Brahmaputra-Meghna
river basin, which drains more than 1.7 million square kilometers
from the Himalaya\textquoteright s headwaters to the Ganges delta
\citep{Frenken12}. While the major areas of the basins are of temperate
climate, the Meghna river basin in Bangladesh is of tropical climate
and is amongst the wettest regions of the world with annual precipitation
that often exceeds 4000 mm \citep{Mirza98}. The ganges basin is heavily
populated by dense irrigated agriculture, especially at the vicinity
of its tributary. Rainfall in the study region is mainly controlled
by the monsoon season from June to September. The Himalayan range
typically forces the southwesterly moist monsoon winds to dump their
moisture over the lowlands and shorelines of the region, while it
leaves the Tibetan plateau and Thar Desert relatively dry in rain
shadows. The deltaic region often experiences very strong cyclonic
storms during the monsoon season, some of which resulted in significant
loss of life and properties. Among these was the cyclone Sidr, hitting
the coastlines of Bangladesh in November 2007.

\subsection{Surface condition and radiation regime}

Figure \ref{fig:1} (top and middle panels) shows the boundary and
important geographic features of the study domain, digital elevation
model, the annual Normalized Difference Vegetation Index (NDVI) and
snow cover percentage in calendar year 2013. The vegetation and snow
cover data are derived by averaging over the L3 global monthly gridded
products at 0.05-degree (MYD13C2, MYD10CM), obtained from the Moderate-resolution
Imaging Spectroradiometer (MODIS) instrument on board the NASA\textquoteright s
Aqua satellite. Figure \ref{fig:1} (bottom panels) also demonstrates
the annual fields of TMI brightness temperatures for the same year
at 22v, 85v channels and their differences. Note that these BT fields
are obtained for precipitation-free scenes, by overlying and averaging
all overpasses of TMI over the inner swath, where the PR-2A25 is used
for eliminating the raining areas. 

\begin{figure}[H]
\noindent \begin{centering}
\includegraphics[width=0.70\paperwidth]{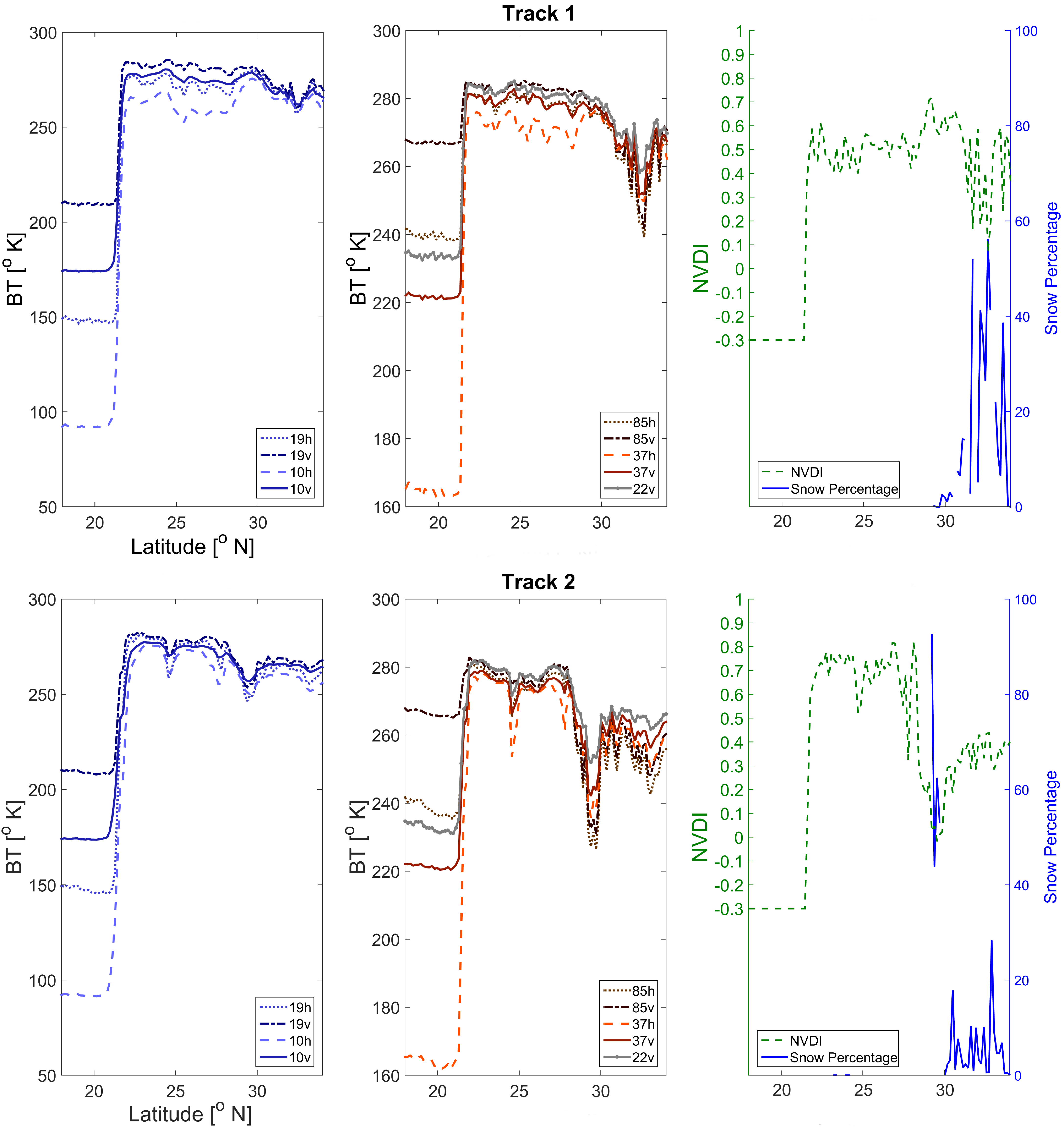}
\par\end{centering}

\protect\caption{1D transects of the non-raining TMI spectral BTs, Normalized Difference
Vegetation Index (NDVI) and annual percentage of snow cover over the
tracks shown in Figure \ref{fig:1}. \label{fig:2}}
\end{figure}

Important radiation patterns can be seen especially over snow-covered
lands, wet surfaces, and at the vicinity of coastlines. In particular,
even though among low-frequency channels the 22v exhibits a minimal
response to the surface snow emissivity, we still see notable depressions
in this channel due to the presence of snow on the ground. More importantly,
we see in the bottom-right panel that the high-frequency depressions
in free-rain scenes, characterized by $\text{SI}={\rm Tb_{22v}^{-}}-{\rm Tb_{85v}^{-}}$, go beyond the prescribed \textasciitilde{}8 Kelvin--especially over
the headwaters of the Brahmaputra Basin and the Hengduan Mountains.
This observation suggests that the SI-method is prone to false rainfall
detection and may gives rise to the rainfall overestimation above
snow-covered lands. Near the coastlines, because of a marked difference
between emissivity of ocean and land surfaces at the low frequency
channels (1-37 GHz), we see a sharp temperature gradient at the interface
while this effect is less pronounced at 85 GHz. It will be shown later
on that this transient radiation regime poses challenges for rainfall
retrieval at the vicinity of coastlines. 

To provide deeper insight into the information content of the low-frequency
channels, relevant to the rainfall retrieval problem, Figure \ref{fig:2}
demonstrates 1D-transects of all frequency channels along the tracks
shown in the digital elevation model shown in Figure \ref{fig:1}.
On one hand, below $28^{\circ}$N, the left track (Track 1) mainly
passes through moderately vegetated lowlands with NDVI=$0.4-0.6$, while
between $20-30^{\circ}$N, it crosses slightly denser vegetation over
the southward slope of the Himalayan range. Over higher latitudes
this track records lower NDVI =$0.1-0.4$ and up to 60\% of annual snow
cover. On the other hand, the right track (Track 2) crosses the Chin
Hill mountain range with tropical and subtropical forest ecosystem
with high NDVI values ranging mostly from 0.6 to 0.8 below latitude
$27^{\circ}$N. For higher latitudes around $20-30^{\circ}$N, this
track goes over highly snow covered and perhaps frozen grounds while
the NDVI index reduces to $0.1-0.4$, indicating the presence of sparse
vegetation and steppe grasslands. In Figure \ref{fig:2}, over the
tributaries of the Ganges river basin below $30^{\circ}$ N latitudes,
Track 1 represents large polarization and frequent radiometrically-cold
spots, especially for frequencies $\leq37$ GHz. For higher latitudes
in this track, we see that the polarization gaps start to shrink over
denser vegetation covers. It can be seen that the snow covered surfaces
create a sharp decrease in BTs and slightly increase polarization
effects, especially for frequency channels $\geq37$ GHz. Note that
while over vegetated areas, channel 85 GHz is typically warmer than
37 GHz, the presence of snow makes it cooler than 37 GHz. In Track
2, we see less polarization, compared to the Track 1, because of denser
vegetation and rougher pathway over the Chin Hills. However, we see
a few significant isolated drops in BT values (see, 10h and 37h) due
to the radiometrically cold wet surfaces, especially where the track
crosses the Meghna-Brahmaputra river system. In this track, a significant
temperature drop can be seen due to the high percentage of snow cover
around $30^{\circ}$N in frequencies greater than 37 GHz. 

\begin{figure}[H]
\noindent \begin{centering}
\includegraphics[width=0.45\paperwidth]{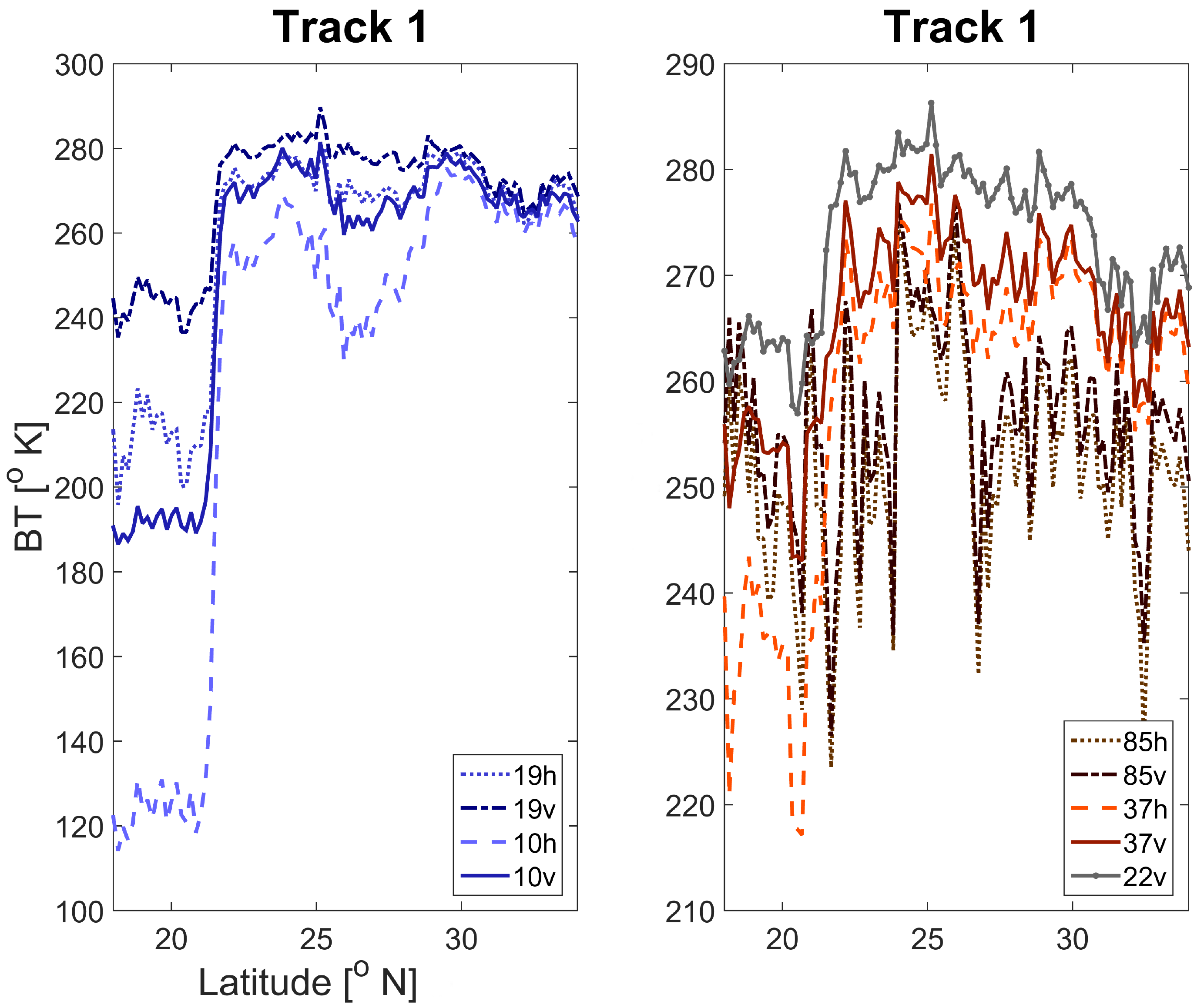}
\par\end{centering}

\protect\caption{1D transects of the TMI raining spectral BTs, over Track 1 shown in
Figure \ref{fig:1} and \ref{fig:2}.\label{fig:3}}

\end{figure}

The annual climatology of the raining BTs along the first track is
shown in Figure \ref{fig:3}. This transect of BT values is obtained
analogous to those explained in Figure \ref{fig:2}, except for the
scenes which are labeled as raining by the PR-2A25 data. The main
goal is to demonstrate the effect of the rainfall signal on high-
and low-frequency channels and provide some intuition that the nearest
neighbor isolation of spectral candidates can provide partial regionalization
skills to the ShARP retrievals. Over land, we clearly see that the
rainfall spectral signals exhibit minimal influence on the low frequency
channels below 19 GHz, although some increased polarization can be
seen compared to non-raining BTs. However, the rainfall signatures
are very pronounced over frequencies greater than 22 GHz and dramatically
distort the background surface radiation pattern.

\begin{figure}[H]
\noindent \begin{centering}
\includegraphics[width=0.75\paperwidth]{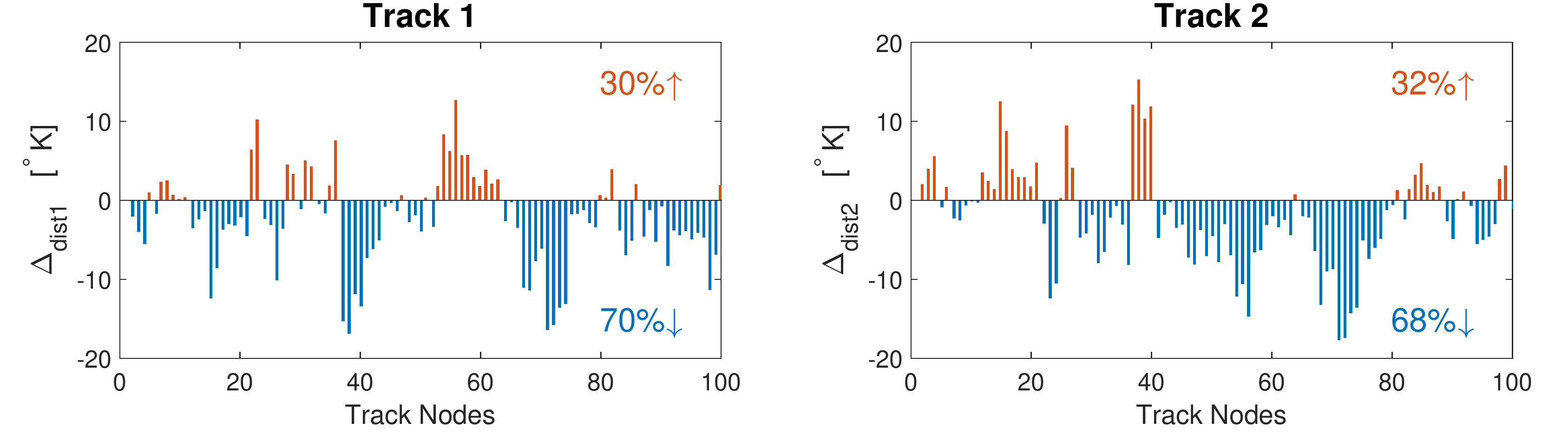}
\par\end{centering}

\protect\caption{Difference in Euclidean distance between the annual non-raining brightness
temperatures along each track with the raining brightness temperatures
along both tracks in calendar year 2013. The bars below the x-axis
show that in almost 70\% of the nodes the raining BTs are closer to
their non-raining counterparts along the same track, which partly
represent the land surface condition. Geographic positions of Track 1 (left panel) and Track
2 (right panel) are shown in Figure \ref{fig:1}. \label{fig:4}}

\end{figure}

Let us assume that at each node of the $i^{th}$ track, here $i=\left\{ 1,\,2\right\} $,
the annual non-raining and raining BTs are represented by $\mathbf{y}_{i}^{-}=\left\{ {\rm Tb_{v}^{-}}\right\} _{{\rm v}\in\mathcal{V}}$
and $\mathbf{y}_{i}^{+}=\left\{ {\rm Tb_{v}^{+}}\right\} _{{\rm v}\in\mathcal{V}}$,
respectively, where $\mathcal{V}$ is the set of available TMI channels.
Furthermore, let us define the Euclidean distance between non-raining
and raining BTs along those tracks as $\text{dist}_{i,j}=\left\Vert \mathbf{y}_{i}^{-}-\mathbf{y}_{j}^{+}\right\Vert _{2}$,
where $\left\Vert \mathbf{y}\right\Vert _{2}^{2}=\sum_{{\rm v}\in\mathcal{V}}{\rm Tb}_{{\rm v}}^{2}$.
Figure \ref{fig:4} shows $\Delta_{{\rm dist1}}={\rm dist}_{1,2}-{\rm dist}_{1,1}$(left
panel) and $\Delta_{{\rm dist2}}={\rm dist}_{2,1}-{\rm dist}_{2,2}$
(right panel) for calendar year 2013. In words, for each track, this
measure of distance-difference characterizes the difference between
the Euclidean distances of its non-raining background signal with
the raining ones of both tracks. We see that the defined measure of
distance-difference falls below the x-axis in almost 70 percent of
the nodes in both tracks. In effect, in 70 percent of cases the raining
nodes along each track are closer to their non-raining counterparts
along the same track, which partly represent the underlying surface
properties. In other words, it is likely that the surface properties
of the nearest spectral neighbors are consistent with the surface
properties under the retrieval scene, an important characteristic
that needs to be taken into account in modern land retrieval techniques
\citep[see,][]{Turk2013a}. By visual inspection and comparison of
Figure \ref{fig:2} and \ref{fig:3}, we can see that this skill mainly
comes from the low-frequency channels (10-19 GHz), for which the observed
BTs mainly respond to the land surface emission and remain almost
insensitive to the atmospheric signals. Thus, one might interpret
this observation as a regionalization skill of the ShARP algorithm. 

\subsection{ShARP parameterizations and settings}

\begin{figure}[H]
\noindent \begin{centering}
\includegraphics[width=0.75\paperwidth]{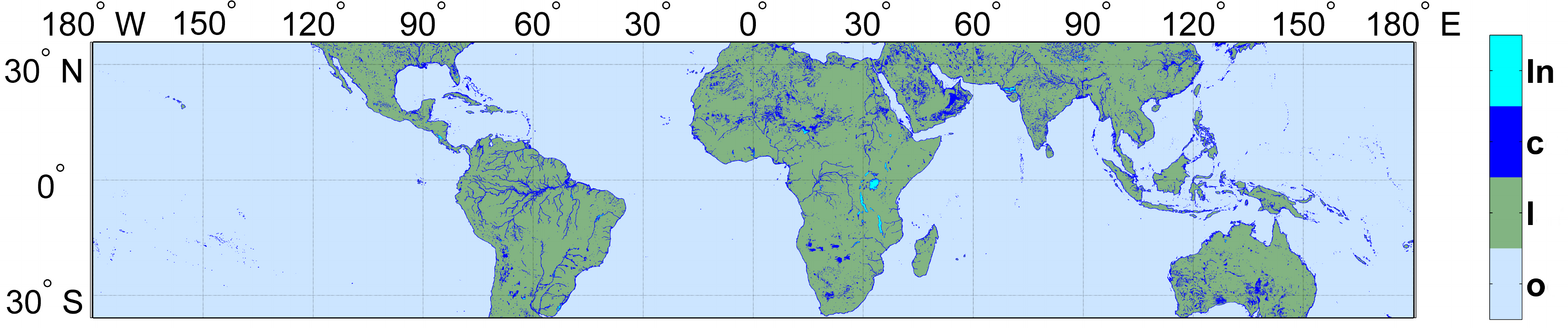}
\par\end{centering}

\protect\caption{ShARP land surface classes include inland water body (In), coastal
zone (c), land (l), and ocean (o). The data is obtained from PR-1C21
product (version 7), mapped onto a regular grid at 0.05-degree \citep{EbtBF15}.\label{fig:5}}

\end{figure}

In this study we run ShARP for four different earth surface classes,
namely: ocean, land, coast, and inland water bodies (see, Figure \ref{fig:5}).
In other words, we collect four dictionaries and run the algorithm
over each surface class depending on the geolocation of each pixel.
This surface stratification is obtained from standard surface data
in the PR-1C21 product (version 7) at \textasciitilde{}5 $\times$
5 km grid-box. To construct spectral and rainfall dictionaries, we
used collocated pixels of the TMI and PR in 2000 orbits, randomly
chosen from a rainfall database collected in five years 2002, 2005,
2008, 2011 and 2012. In these sampled orbits, more than 25 million
raining and non-raining pixels were used to construct the required
dictionaries. To obtain collocated radar-radiometer data, we focused
on the swath-level calibrated BT values and near surface rainfall
in TMI-1B11 and PR-2A25 \citep{IguKMA00} products, respectively.
To this end, we mapped the TMI measurements onto the PR grids using
the nearest neighbor interpolation. 

In the detection step, we found that the probabilities of hit and
false alarm are not very sensitive to the number of nearest neighbors
for $k\geq10$ and reasonable detection accuracy can be obtained by
assuming $10\leq k\leq50$. Notice that in the detection step, we
consider the information content of each channel equally important
for the nearest neighborhood search. However, in the estimation step,
we use a weight or precision matrix $\mathbf{W}$ in problem (\ref{eq:2})
to properly encode the relative importance of each channel for rainfall
estimation. For instance, we assign larger weights to low-frequency
channels (e.g., 10 GHz) for retrievals over ocean while high-frequency
channels (e.g., 85 GHz) receive larger weights over land. To this
end, we used a diagonal precision matrix for which the diagonal elements
are filled with the coefficients of variation of raining BTs in each
channel, obtained from the collected dictionaries \citep[see,][]{EbtBF15}. 

\begin{figure}[H]
\noindent \begin{centering}
\includegraphics[width=0.75\paperwidth]{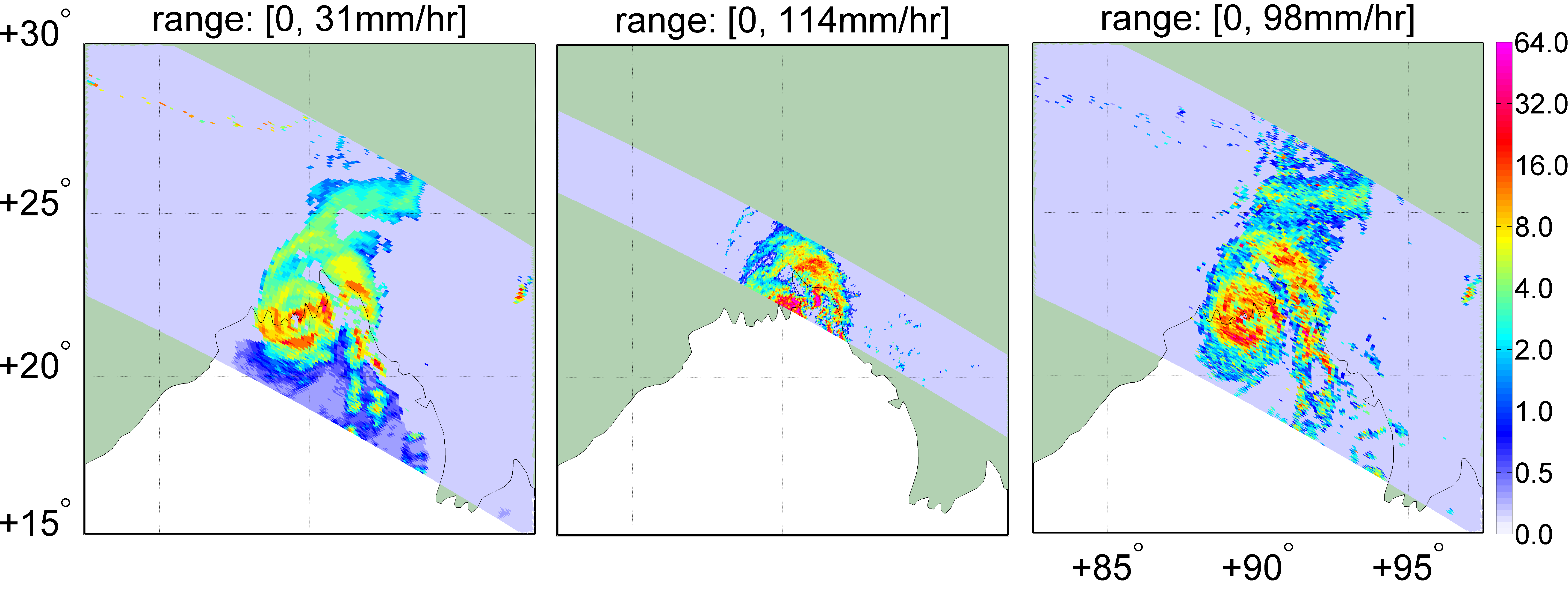}
\par\end{centering}

\protect\caption{Rainfall retrievals for the TRMM overpass on November 15, 2007 capturing
the cyclone Sidr at UTC 13:59, including the results from: 2A12 (left panel), 2A25 (middle panel), and ShARP (right
panel).\label{fig:6}}
\end{figure}

\subsection{Instantaneous Retrieval of the Cyclone Sidr}

Cyclone Sidr has been one of the most intense and historically disastrous
tropical cyclones that hit the coastline of Bangladesh on November
15, 2007. TRMM satellite flew over the cyclone on November 15 at 13:59
UTC and provided critical information about the storm vertical structure
and spatial extent. Figure \ref{fig:6} shows the retrieved rainfall
rates obtained from ShARP and compares it with the standard version
7 of the PR-2A25 and TMI-2A12.

\begin{figure}[H]
\noindent \begin{centering}
\includegraphics[width=0.70\paperwidth]{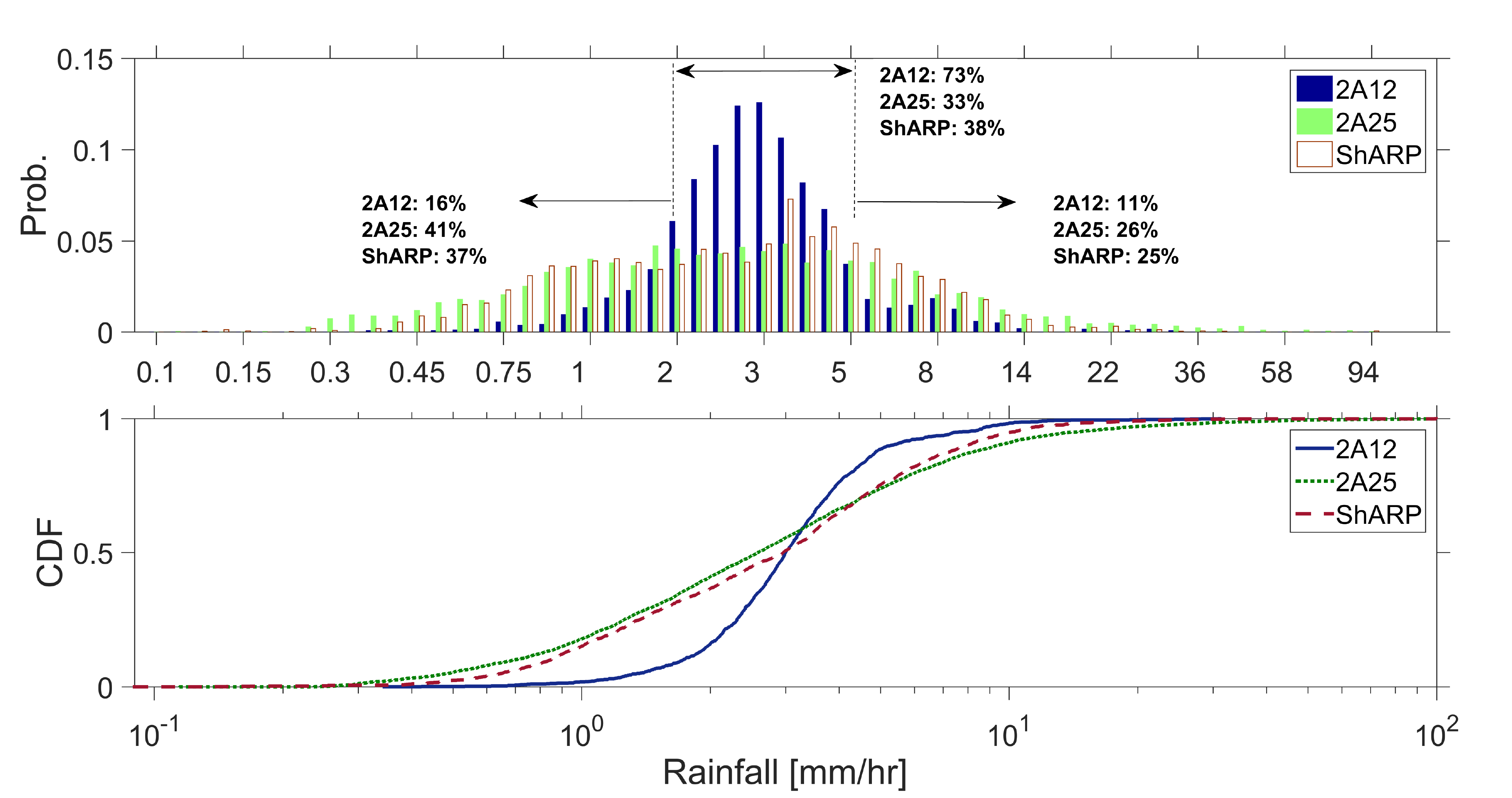}
\par\end{centering}

\protect\caption{Pixel level probability histograms (top) and empirical cumulative
density function (bottom) of the retrieved rainfall over the inner swath
of the TRMM overpass, capturing the cyclone Sidr on November 15, 2007
at 13:59 UTC. \label{fig:7}}
\end{figure}

We see that the multiband structure of the cyclone has been well retrieved
by ShARP (right panel). Comparing the results with 2A25 (middle panel)
and 2A12 (left panel), it is evident that the ShARP retrieval properly
captures the high intense rainbands of the cyclone over land and at
the vicinity of the coastline. For instance, the intense rainfall
band--right above the Ganges deltas--is estimated around 8 mm/hr in
the 2A12 while this estimate in both ShARP and 2A25 is around 16 mm/hr.
Furthermore, at the vicinity of the coastlines, we see some discontinuities
over ocean in the 2A12 product. In these regions, it can be seen that
some of the light rain patches close to the shoreline are underestimated
or completely missed. However, the ShARP retrievals remain relatively
coherent close to the shoreline and some isolated rainfall patches
are also recovered. 

In light of lacking an absolute reference rainfall field for the above
snapshot, objective comparison of these retrievals is not straightforward,
especially over ocean where the data bases are of different nature
(i.e., observational for ShARP and mainly model-based for 2A12). However,
Figure \ref{fig:7} compares the histogram and empirical cumulative
density function (CDF) of the retrieved rainfall over the inner swath,
which mostly captures the rainfall over land and at the vicinity of
shorelines, where both algorithms attempt to reproduce PR-2A25 rainfall.
We clearly see that ShARP performs well in retrieving a wide range
of rainfall intensities and properly resembles the distribution of
rainfall provided by the reference 2A25. It is seen that ShARP retrieves
rainfall rates below 1 mm/hr and above 14 mm/hr, while a large probability
mass of the retrieved rainfall remains between 2-5 mm/hr in the 2A12.
The dynamic range of the retrieved rainfalls is better illustrated
in the CDF plots. Consistent with the shown histograms, we see that
the passively retrieved rainfall values are mostly concentrated around
the mode in 2A12 while the results in ShARP are sufficiently stretched
and properly resemble the reference 2A25 rainfalls. Obviously, both
retrievals fall short to properly explain the thick tail distribution
of the PR-derived rainfall, however, this problem seems to be ameliorated
in ShARP.

The proximity of the shown rainfall histograms might be measured by
the Kullback-Leibler (KL) divergence. The KL is a non-negative measure
that quantifies the degree of closeness between a reference probability
distribution $\mathcal{P}=\left\{ p_{i}\right\} $ and an approximate
one $\mathcal{P}=\left\{ \hat{p}_{i}\right\} $, 

\begin{equation}
KL\left(\mathcal{P}||\hat{\mathcal{P}}\right)=\sum_{i}p_{i}\log\frac{p_{i}}{\hat{p}_{i}}.
\end{equation}

As is evident, this measure is not symmetric and is zero when the
distributions are identical. A simple calculation shows that the distribution
of ShARP is much closer to the distribution of the 2A25 than the 2A12,
as an estimate of this measure reduces from 0.6 (2A12 vs 2A25) to
0.06 (ShARP vs 2A25).

\subsection{Monthly and annual retrievals }

To better quantify the quality of ShARP retrievals, we compare its
monthly and annual retrievals with the standard 2A25 and 2A12 over
the study domain in calendar year 2013. Even though we show the retrieval
results both over ocean and land, we confine our interpretation only
to the results over land and vicinity of coastlines, where both retrievals
are empirical in their nature and attempt to reproduce the 2A25. We
need to once again emphasize that 2A25 is not free of error; however,
it provides one of the best estimate of the total rainfall at a global
scale \citep[e.g.,][]{Berg2006}.

\begin{figure}[t]
\noindent \begin{centering}
\includegraphics[width=0.65\paperwidth]{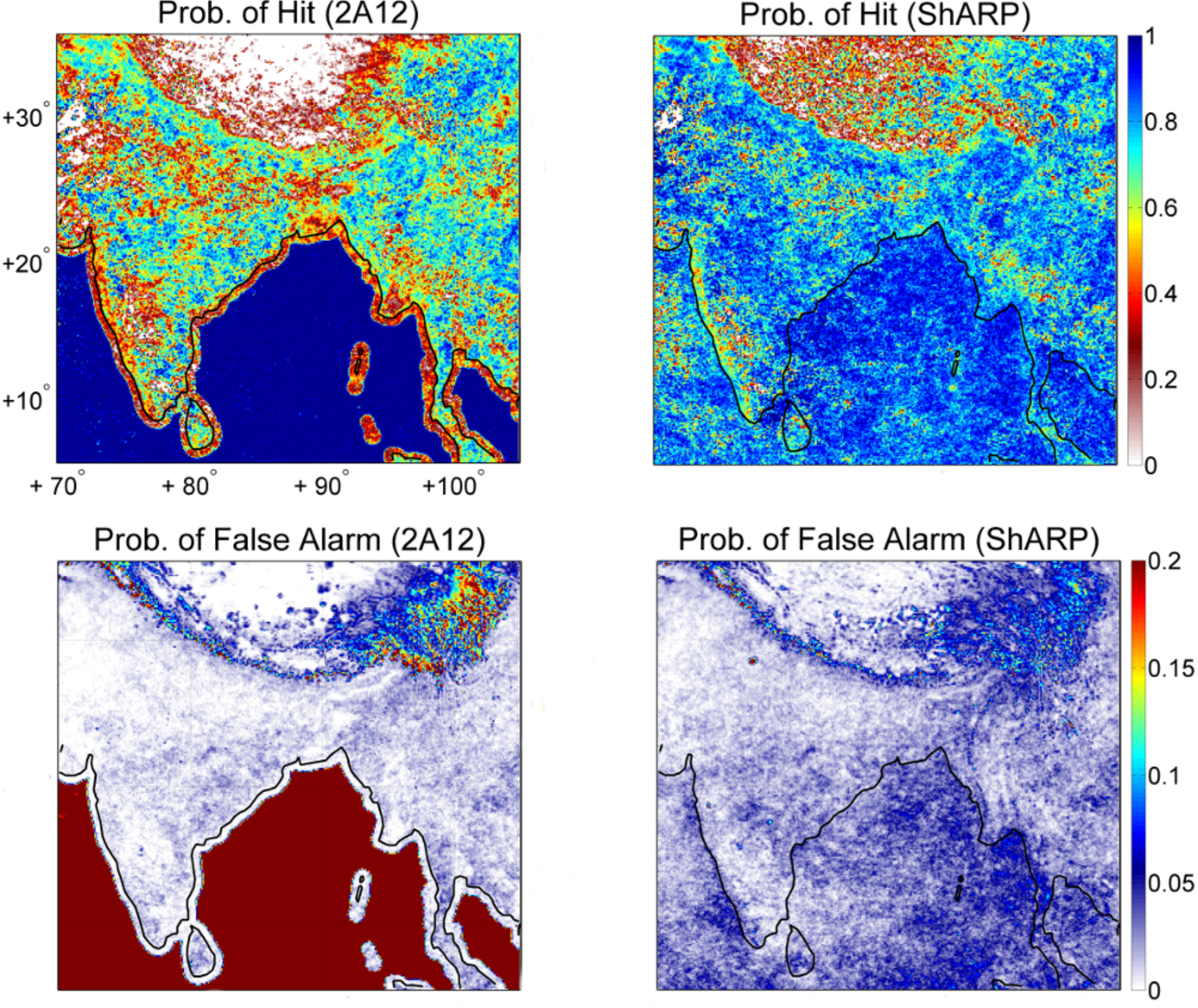}
\par\end{centering}

\protect\caption{Annual probability of hit and false alarm, comparing the 2A12 and
ShARP retrievals with the 2A25 as a reference. The fields show probabilities
obtained for all inner-swath overpasses in 2013 at 0.1-degree resolution.
\label{fig:8}}

\end{figure}
To this end, We first focus on the detection capabilities of ShARP,
especially at the vicinity of coastlines and snow-covered areas. Figure
\ref{fig:8} shows maps of probabilities of hit and false alarm while
the retrieval results of ShARP and 2A12 are binary-wised compared
with the 2A25 for all inner-swath overpasses in 2013. We can clearly
see two main points in ShARP retrievals: 1) coherent and improved
detection rate near the coastlines, and 2) improvement of rainfall
detection over snow-covered land surfaces above the Himalayan range,
headwaters of the Brahmaputra Basin and Hengduan Mountains (Figure
\ref{fig:1}). This figure illustrates that the 2A12 is prone to detect
fewer raining events than the 2A25 near the coastlines. Indeed, we
see a systematic decrease in the probability of detection over a band
of approximately 25-35 km parallel to the coastlines in 2A12. In this
band, the probability of hit mostly falls between 0.2-0.5 in 2A12
while it is within the range of 0.7-1.0 in ShARP. Moreover, over the
tributaries of the Ganges river basin and its delta, with dense agricultural
activities near the river banks and thus high soil moisture content,
we observe improved probability of hit in ShARP. For the strip of
Malabar Coast with tropical monsoon climate--between the Western Ghats
mountain range and southwest coastlines of the Indian peninsula--we
also see that the probability of hit in both retrievals decreases
appreciably. However, ShARP shows some improvements in this regime
as well. Note that in this wet strip the rainfall processes are mainly
orographic and highly seasonal whose annual amount is mostly governed
by the monsoon season from May to September. Therefore, we suspect
that the decrease in probability of hit mostly corresponds to the
shallow rainfall regimes, when and where the ice signature is week
on high-frequency channels and masked by the sharp gradient of the
background emissions. We also observe that both methods show low detection
skills in capturing sporadic and low-intensity rainfall over the dry
region of the Tibetan plateau. Specifically, the ShARP detection rate
over the northern part of the plateau falls between 0.0-0.5, while
this rate tends to zero in 2A12. However, we will demonstrate in the
sequel that these low values correspond to light rainfall amounts
that may not exceed 5-20 mm per year. The maps of the probability
of false alarm (Figure \ref{fig:8}, bottom rows) reveal advantages
of ShARP in retrieving rainfall over snow-covered land. In particular,
notable decrease in the probability of false alarm over snow-covered
surfaces in the Himalayan range and Hengduan Mountains is seen in
ShARP retrievals.

\begin{figure}
\noindent \begin{centering}
\includegraphics[width=0.80\paperwidth]{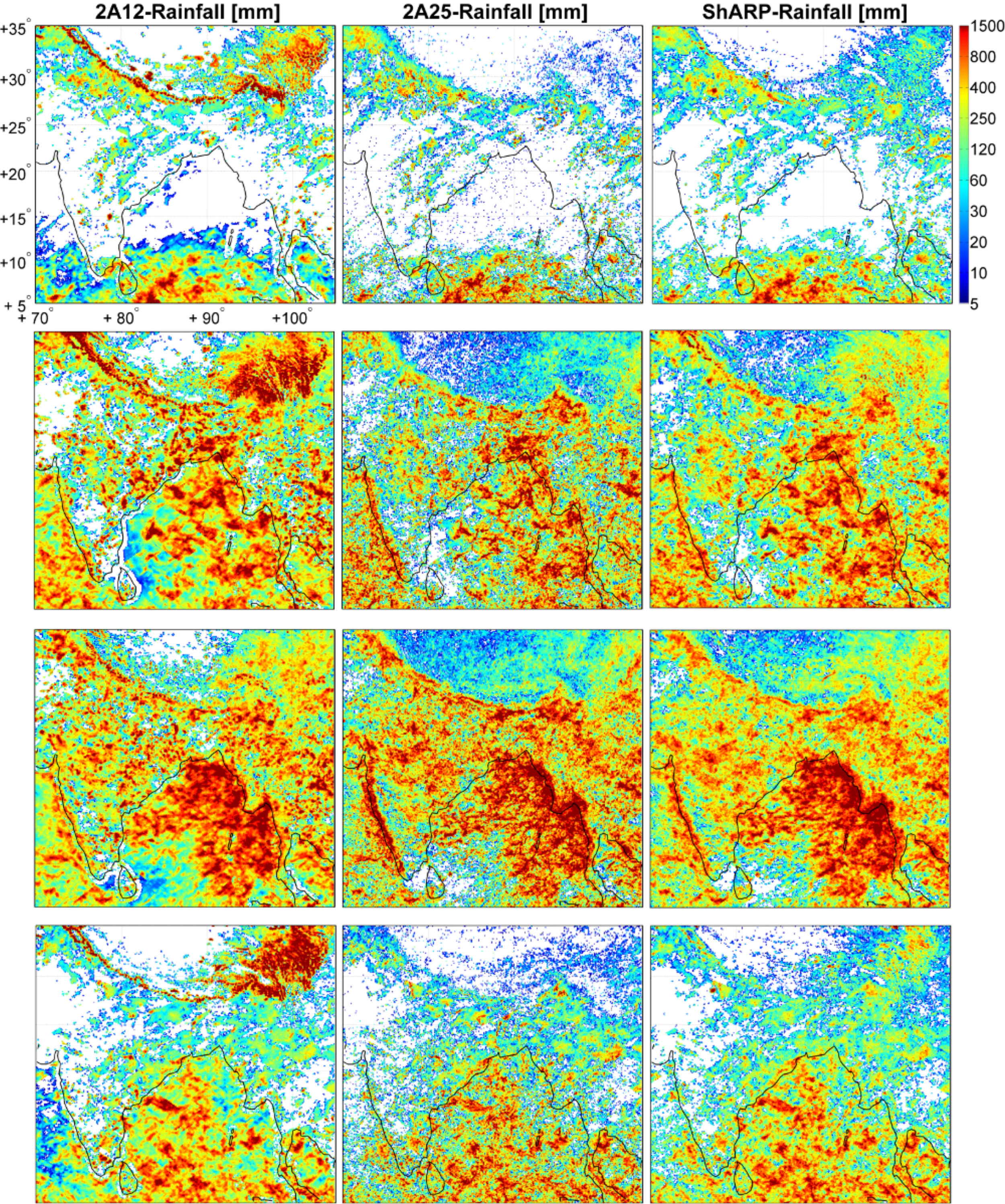}
\par\end{centering}

\protect\caption{Monthly rainfall retrievals over the study region shown at 0.1-degree.
First to the last rows contain three months accumulations of total
rainfall (mm) for January-March (JFM); April-June (AMJ); July-September
(JAS) and October-December (OND) throughout the calendar year 2013.
The results only contain the rainfall captured within the TRMM inner
swath.\label{fig:9}}
\end{figure}

Figure \ref{fig:9} compares the seasonal retrievals of ShARP with
the TRMM standard products. The objective is to further demonstrate
the performance of ShARP retrievals close to the shorelines and over
snow-covered mountainous regions. Indeed each row in Figure \ref{fig:9},
shows three-month rainfall {[}mm{]} from January to December in 2013.
During the cold seasons such as January-March (JFM, first row in Figure
\ref{fig:7}) and October-December (OND, last row in Figure \ref{fig:9}),
we see marked over estimation in passive retrievals along the Himalayan
range and southeast of the plateau--covering the Hengduan Mountains.
However, as is evident, this over estimation is markedly reduced in
the ShARP retrieval. This improvement is largely due to the multispectral
nature of ShARP and its nearest neighbor screening in the detection
step. Indeed, as explained, information content of the lower frequency
channels allows us to better filter out the noisy background signal
of snow and constrain the estimation step only to a few physically
relevant candidates in the dictionaries. It is worth nothing that,
this improvement persists during the snow melting season--April-June
(AMJ, second row in Figure \ref{fig:9})--indicating the robustness
of our algorithm against the drastic changes in the temporal dynamics
of snow emissivity.

During the warm months of the monsoon season in July-September (JAS,
third row in Figure \ref{fig:9}), we see that both retrieval methods
reproduce well the target 2A25 over the Tibetan highlands and Himalayas.
However, consistent with the observations in Figure \ref{fig:8},
some rainfall underestimation can be spotted over the coastal strip
of the Malabar region in southeast India and the Rakhine coasts in
the western Myanmar. The rainfall regime over both of these regions
is heavily influenced by the presence of the Western Ghats and Arakan
orographic barriers blocking the moist southwesterly monsoon winds.
Among these two coastal areas, it is important to note that the land-ocean
interfacial radiation regime seems to be more complex at the vicinity
of the Rakhine coastlines, particularly due to the presence of complex
deltaic landforms and the dynamics of multiple river mouths. The observed
underestimation in passive retrievals over the strip of the Malabar,
seems to be pronounced both over coasts and shores, perhaps indicating
that the total rainfall is largely controlled by prolonged periods
of warm and shallow orographically enhanced events. While less significant
under estimation, over the coasts of the Rakhine State, might be an
indication that the rainfall is more influenced by deeper convections
and more cold clouds in this region, compared with the Malabar strip. 

\begin{figure}
\noindent \begin{centering}
\includegraphics[width=0.75\paperwidth]{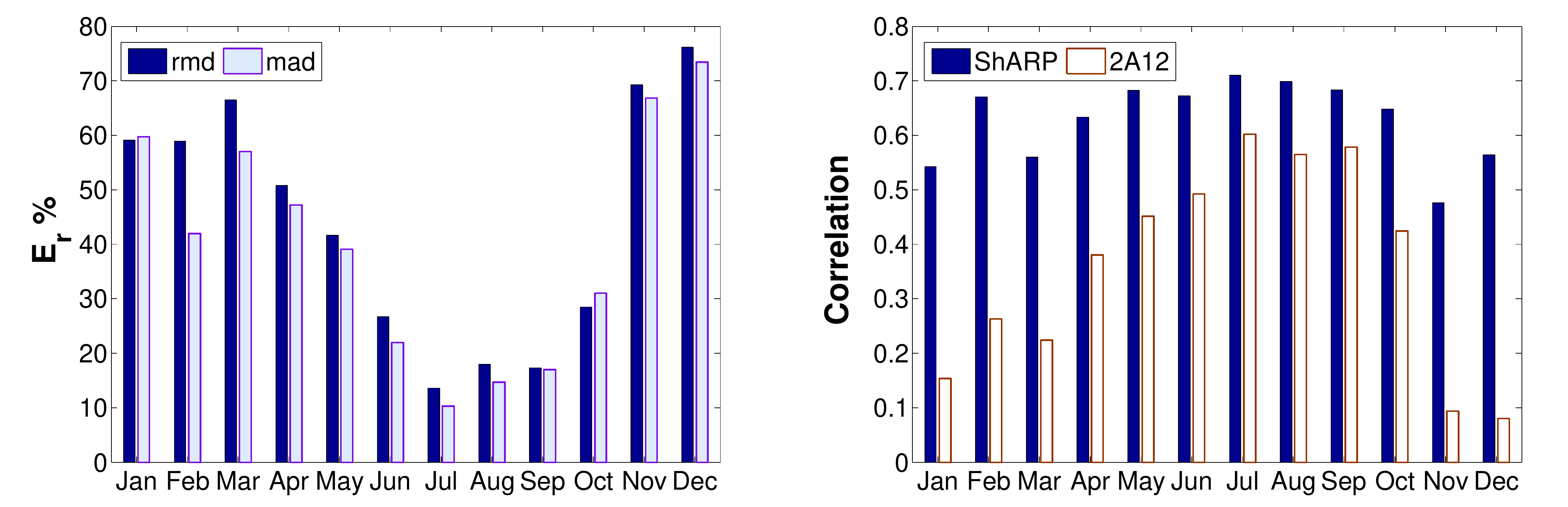}
\par\end{centering}

\protect\caption{Left panel: relative reduction ($\mathbf{E}_{{\rm r}}$\%) in monthly
root mean-squared difference (rmd) and mean absolute difference (mad)
in ShARP retrievals compared with the 2A12. The results are confined
to land and coastal areas of the study domain in 2013. The reference
rainfall is set to be 2A25. Right panel: the correlation coefficients
of the monthly retrieved fields with the 2A25.\label{fig:10}}

\end{figure}

\begin{figure}[H]
\noindent \begin{centering}
\includegraphics[width=0.75\paperwidth]{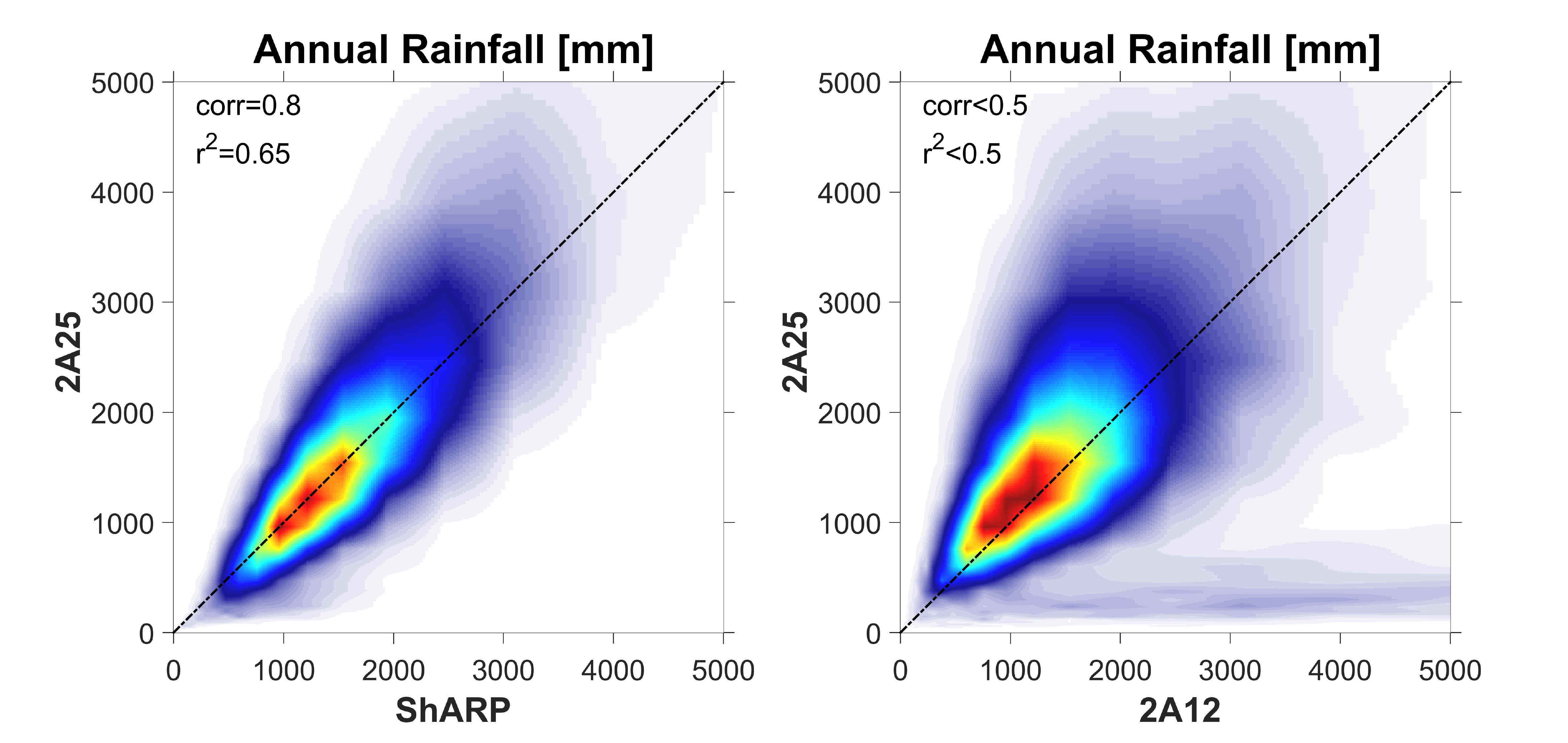}
\par\end{centering}

\protect\caption{Smooth scatter plots of the total rainfall {[}mm{]} for PR-2A25 versus
ShARP (left panel) and 2A12 (right panel) in 2013. The plots are obtained
from pairs of the total rainfall over land and coastal areas of the
study domain projected onto a regular grid at 0.2-degree.\label{fig:11}}

\end{figure}

Figure \ref{fig:10} shows some quantitative measures that compare
monthly retrievals in 2013, restricted to land and coastal areas of
the study domain. Specifically, we computed the monthly root mean
squared difference (rmd) and mean absolute difference (mad) of both
retrievals with the 2A25 and reported their relative reductions, e.g.,
(rmd\textsubscript{ShARP} \textendash{} rmd\textsubscript{2A12})/rmd\textsubscript{2A12}
(Figure \ref{fig:10}, left panel). As is evident, ShARP is markedly
closer to the 2A25 throughout the entire year, especially over cold
seasons. The reduction in the chosen proximity metrics reaches up
to 75\% in snow falling months of November and October. As expected,
these metrics reach their minimum values (10-15\%) during the warm
months of July-September, over which, retrieval is less challenging
for the SI-based techniques. We see notable differences between the
reduction of rmd and mad metrics over the dry JFM period. Note that,
the rmd quadratically penalizes the error and thus is more sensitive
to large values and the tail of the distribution of retrieval differences
than the mad metric. Thus, during the JFM period, the distribution
of retrieval differences seems to be populated by large over estimations
and thus shall be heavier tailed compared to the other months. The
correlations (Figure \ref{fig:10}, right panel) of the monthly retrievals
with the 2A25 also confirm the observed trend in the evaluated proximity
metrics. Indeed, we see that the quality of the monthly ShARP retrievals
is robust and remains relatively independent of the seasonal rainfall
variability and land surface radiation regime change.

Finally, density scatter plots of the retrieved annual rainfall are
shown in Figure \ref{fig:11}. The results in this figure demonstrate
pixel-level pairs of the annual rainfall estimates over land, in 2013,
projected onto a regular grid at 0.2-degree. The density scatter plots
confirm that both passive retrievals are slightly subject to underestimation
of the high-intense rainfall patches. We can also see that an appreciable
amount of rainfall pairs are away from the 1:1 line in the 2A12, pointing
to the notable overestimation reported in Figure \ref{fig:9}. Comparing
annual passive retrievals with the 2A25, the annual rmd in ShARP is
38\% smaller than the one in the 2A12, mainly because of improved
retrievals of rainfall over snow-covered ground. Note that, due to
the reported over estimation, the annual correlation (corr) and coefficients
of determination ($\mathbf{r}^{2}$) in 2A12 are less than 0.5, while
these statistics reach 0.8 and 0.65 in ShARP, respectively.

\section{Concluding Remarks\label{sec:4}}

We presented promising results for passive rainfall retrieval over
land and coastal zones, using the recently proposed Shrunken locally
linear embedding Algorithm for Retrieval of Precipitation \citep[ShARP,][]{EbtBF15}.
We provided evidence that as ShARP uses information across all spectral
channels, it allows to properly discriminate background radiation
from the rainfall signal, especially over frozen and snow-covered
grounds. In addition, at the vicinity of coastlines, some improved
results in rainfall detection and estimation are also elucidated which
promise a step forward for applications related to snow and coastal
hydrology in the era of the Global Precipitation Measuring (GPM) project. 

Research is currently underway to extend ShARP to take explicitly
into account the ground surface spectral emissivity as well as databases
from multiple platforms and satellites with different spectral and
spatial resolutions. While ShARP has been tested for observationally-based
TRMM data, its extension to use combined physically-based and observationally-generated
dictionaries might be of future research interest.

\section*{Acknowledgment}

The authors gratefully acknowledge the financial support provided
by the K. Harrison Brown Family Chair funding during the course of
this research. Furthermore, the support provided by two NASA Global
Precipitation Measurement grants (NNX13AG33G, NNX13AH35G) are also
greatly recognized. The TRMM 2A12 and 2A25 data were obtained through
the anonymous File Transfer Protocol publicly available at \url{ftp://trmmopen.gsfc.nasa.gov/pub/trmmdata}.
The MODIS retrievals (i.e., MYD13C2 and MYD10CM) were acquired
from the Land Processes Distributed Active Archive Center (LP DAAC), publicly accessible at \url{https://lpdaac.usgs.gov/data_access/data_pool}.

\bibliographystyle{agufull04}

\end{document}